\documentclass[fleqn,usenatbib]{mnras}
\pdfoutput=1
\usepackage{newtxtext,newtxmath}

\usepackage[T1]{fontenc}
\usepackage{ae,aecompl}

\usepackage{graphicx}	
\usepackage{amsmath}	
\usepackage{amssymb}	

\usepackage{epsfig}
\usepackage{rotating}
\usepackage{multirow}
\usepackage{longtable}
\usepackage{breqn}
\usepackage{amssymb}

\title[Sub-arcsecond Imaging of 0313$-$192]{The First VLBI Detection of a Spiral DRAGN Core}

\author[M. Y. Mao et al.]{Minnie Y. Mao,$^{1,3}$\thanks{E-mail: minnieyuanmao@gmail.com}
Jay M. Blanchard,$^{2}$
Frazer Owen,$^{3}$
Lor\'{a}nt O. Sjouwerman,$^{3}$
\newauthor
Vikram Singh,$^{2,4}$
Anna Scaife,$^{1}$
Zsolt Paragi,$^{2}$
Ray P. Norris,$^{5,6}$
Emmanuel Momjian,$^{3}$
\newauthor
Gia Johnson,$^{3,7}$ and
Ian Browne,$^{1}$ 
\\\\
$^{1}$Jodrell Bank Centre for Astrophysics, Alan Turing Building, School of Physics and Astronomy, The University of Manchester, 
\\Oxford Road, Manchester, M13 9PL, UK\\
$^{2}$Joint Institute for VLBI ERIC (JIVE), Postbus 2, 7990 AA, Dwingeloo, The Netherlands\\
$^{3}$National Radio Astronomy Observatory, PO Box O, Socorro, NM 87801, USA\\
$^{4}$University of Wyoming, 1000 E University Ave, Laramie, WY 82071, USA\\
$^{5}$Western Sydney University, Locked Bag 1797, Penrith South, NSW 1797, Australia\\
$^{6}$CSIRO Astronomy and Space Science, PO Box 76, Epping, NSW, 1710, Australia\\
$^{7}$University of Arizona, Tucson, AZ 85721, USA\\
}

\date{Accepted 2018 May 01. Received 2018 May 01; in original form 2017 October 29}

\pubyear{2017}
\hypersetup{final}
\begin{document}
\label{firstpage}
\pagerange{\pageref{firstpage}--\pageref{lastpage}}
\maketitle

\begin{abstract}
We present the first observation of 0313$-$192, the archetypal spiral DRAGN, at VLBI resolutions. Spiral  DRAGNs are Double Radio Sources Associated with Galactic Nuclei (DRAGNs) that are hosted by spiral galaxies. 0313$-$192 is an edge-on spiral galaxy that appears to host a 360\,kpc double-lobed radio source. The core of this galaxy is clearly detected at L, S, and X-bands using the VLBA, signifying an ongoing active nucleus in the galaxy. This rules out the possibility that the spiral DRAGN is merely a chance alignment. The radio core has L$_{1.4\,GHz} \sim 3.0 \times 10^{23}$\,W\,Hz$^{-1}$. Radio components are detected to the South-West of the core, but there are no detections of a counterjet. Assuming a symmetric, relativistic jet, we estimate an upper limit to the inclination angle of $\theta \lesssim 72$ degrees. The VLBI-detected radio jet components are extremely well-aligned with the larger-scale radio source suggesting little to no jet disruption or interaction with the ISM of the host galaxy. 
\end{abstract}

\begin{keywords}
galaxies: active < Galaxies, galaxies: general < Galaxies, galaxies: jets <
Galaxies, galaxies: spiral < Galaxies, radio continuum: galaxies < Resolved
and unresolved sources as a function of wavelength
\end{keywords}



\section{Introduction}

Spiral DRAGNs are spiral galaxies that host Double-lobed Radio sources Associated with Galactic Nuclei \citep[DRAGN, ][]{Leahy93}. They are not predicted by standard galaxy formation models \citep[e.g.][]{Hopkins08}, which require a major merger to trigger a DRAGN \citep{Chiaberge15}, yet such mergers would destroy spiral morphology. There are instances where radio-loud AGN are hosted by spiral galaxies, as in Seyfert galaxies. However, these radio structures are generally sub-kpc and confined within the scale-height of the host galaxy \citep[e.g.][]{Ulvestad89}. Accretion of matter onto the central black hole can produce a radio-loud AGN, however models suggest that a major merger is required to launch kpc-scale jets \citep{Chiaberge11}. Empirically, DRAGNs in the local Universe are almost always hosted by elliptical galaxies.  

The archetypal Spiral DRAGN 0313$-$192 (WISE J031552.09-190644.2, z = 0.067) was first reported by \citet{Ledlow98,Ledlow01}. Follow-up observations by \citet{Keel06} with the Hubble Space Telescope suggest the host galaxy to be an edge-on spiral galaxy, likely Hubble type Sb. This is the first instance of a spiral galaxy hosting a large-scale double-lobed radio source. Some Seyfert galaxies are hosted by spiral galaxies, however, the radio structures associated with Seyfert galaxies never exceed the scale-height of the host galaxy. All examples of spiral DRAGNs to-date host DRAGNs that are at least an order of magnitude larger in physical size than the host galaxy \citep[e.g.][]{Mao15, Ledlow01, Mulcahy16}.

Since the discovery of 0313$-$192, there have been a number of spiral DRAGNs reported in the literature \citep[e.g][]{Hota11, Bagchi14,Mao15,Singh15,Mulcahy16}. \citet{Mao15} performed a systematic search for these enigmatic sources by cross-matching an optically selected sample of spiral galaxies from Galaxy Zoo \citep{Lintott08} with extended radio sources from FIRST and NVSS \citep{Becker95, Condon98} and found only one. The small number of discovered spiral DRAGNs in the Universe suggests that these sources are very rare. Understanding spiral DRAGNs may reveal insights in galaxy formation.

0313$-$192 hosts a 360\,kpc DRAGN with a 1.4\,GHz integrated flux density of 98\,mJy, which at z = 0.067, translates to a total power of L$_{1.4\,\textrm{GHz}} \sim 10^{24}$\,W\,Hz$^{-1}$, typical to the FRI population \citep{Fanaroff74}. The large-scale radio morphology is also consistent with an FRI morphological type \citep[e.g.][]{Ledlow98}. The spiral host galaxy resides in the poor cluster Abell 428. 

The AGN shows no signs of variability, the Parkes PMN survey in 1994 found a flux density of $87 \pm 11$ mJy at 4.8 GHz \citep{Griffith1994}, while a flux density of $99 \pm 5$ mJy was measured at 5 GHz as part of the AT20G survey in 2010 \citep{Murphy2010}. 

\citet{Keel06} suggest that the DRAGN associated with the spiral galaxy 0313$-$192 may be due to the combination of a number of factors. Firstly, the bulge of the spiral galaxy is overluminous, suggesting that the central black hole is unusually massive thus the associated radio jets may be able to travel faster than, for example, in Seyfert galaxies. Moreover, the radio lobes are almost exactly perpendicular to the disc of the host galaxy in the plane of the sky. The combination of powerful fast radio jets with less interstellar medium for them to burrow through may allow for radio lobes to develop. Finally, the spiral galaxy appears to have a warped stellar disc, which is evidence for a minor merger in the galaxy's past. Given that a merger is currently predicted to be necessary for the formation of DRAGNs, understanding whether a minor merger occurred in the history of this galaxy is vital for constraining formation models of galaxies and DRAGNs.

Interestingly, all of the other spiral DRAGNs detected to date also appear to have unusually massive bulges, and reside in overdense environments. 


The rarity of spiral DRAGNs provides an unprecedented opportunity to challenge the standard galaxy formation model. However, the veracity of these sources must first be confirmed. \citet{Shaver83} presented the first putative spiral DRAGN in the literature. With higher resolution optical data, they determined that the large-scale DRAGN seemingly associated with the host spiral galaxy was in fact a chance-alignment and actually associated with a background elliptical galaxy. In order to confirm the association of the DRAGN with the spiral galaxy 0313$-$192 we observed the source using VLBI. VLBI observations of 0313$-$192 enable us to accurately locate the core of the radio source so we can confirm that it is coincident with the centre of the optical galaxy, rather being a chance superposition. Furthermore, a VLBI detection would confirm that the central engine is active.

This paper presents the results of our VLBI observations of the core of 0313$-$192.



This paper uses H$_0$ = 71 km s$^{-1}$ Mpc$^{-1}$, $\Omega$$_M$ = 0.27 and $\Omega$$_\Lambda$ = 0.73 \citep{Spergel03} and the web-based calculator of \citet{Wright06} to estimate the physical parameters. At the redshift of 0313$-$192 (z\,=\,0.067), one farcsecond corresponds to 1.292\,kpc.

\section{Observations and Data Reduction}

\subsection{LBA}
0313$-$192 was observed at 2.3\,GHz using the standard S-band system of the Australian Long Baseline Array (LBA) in March 2009 (Project code V293). Please see table \ref{obs} for more details. The observation consisted of ten and a half hours of phase referenced observations involving blocks of eight minutes on 0313$-$192 and two minutes on the phase reference source 0325-222. Two ten minute scans on 1921-293 and 0537-441 were used for fringe finding and bandpass calibration. Unfortunately these data had some calibration issues, compromising both the sensitivity and resolution of the data, resulting in a final image with an rms of 71\,$\mu$Jy beam$^{-1}$ and beam FWHM 18.7$\times$15.5\,mas.




\subsection{VLBA}
0313$-$192 was observed at 1.4\,GHz, 2.3\,GHz, and 8.4\,GHz using the standard L, S, and X-band systems of the Very Long Baseline Array (VLBA) in January to August 2013 (Project code BM376). Please see table \ref{obs} for details on frequencies and observations.
Each band was observed for 3 $\times$ 3 hours for a total of 9 hours total observing per band. The observations were dynamically scheduled and the observations covered a range of hour angles. The observations were conducted using nodding-style phase referencing with four minutes on 0313$-$192, and one minute on the phase calibrator, J0315-1656. The strong source J0339-0146 was used as the fringe finder and bandpass calibrator. 


\begin{table}
\begin{center}
\caption{Observation Summary. $\nu$ denotes the central frequency in MHz. V293A was observed with the LBA, the remaining observations were taken with the VLBA. The VLBA S-band observations are limited to 200 MHz usable bandwidth due to RFI tuning filters.}\label{obs}
\begin{tabular}{lllllll}
\\
\hline
Obs ID & Band & $\nu$ (MHz) & $\Delta\nu$ & Date & UT range\\
\hline
V293A  & S & 2301 & 32 & 20090301 & 0200-1230\\
BM376A & X & 8486 & 256 & 20130121 & 0040-0340\\
BM376B & X & 8486 & 256 & 20130304 & 2151-0051\\
BM376C & X & 8486 & 256 & 20130326 & 2024-2324\\
BM376D & S & 2284 & 200 & 20130127 & 0016-0316\\
BM376E & S & 2284 & 200 & 20130426 & 1822-2122\\
BM376F & S & 2284 & 200 & 20130429 & 1811-2111\\
BM376G & L & 1444 & 256 & 20130131 & 2357-0257\\
BM376H & L & 1444 & 256 & 20130818 & 1054-1354\\
BM376I & L & 1444 & 256 & 20130419 & 1850-2150\\
\hline
\end{tabular}
\end{center}
\end{table}

The data were processed using the Astronomical Image Processing System \citep[AIPS, ][]{Greisen03} following the standard data reduction recipe set out in Appendix C of the AIPS Cookbook. The AIPS tasks {\tt uvflg}, {\tt spflg}, and {\tt rflag} were utilised for flagging bad data. Minimal flagging was necessary for L- and X-bands, however the S-band data were severely hampered by RFI and five of the eight IFs were almost completely flagged. LA was used as the reference antenna, with PT used when LA was unavailable. 

Imaging and self-calibration was performed using DIFMAP \citep{Shepherd1994}. 

\begin{table*}
\begin{center}
\caption{Beam size and flux density measurements for the VLBA observations using different imaging weightings. We assume 10$\,\%$ errors, which is standard for VLBI measurements due to the sparse \emph{uv}-coverage.}\label{flux}
\begin{tabular}{lllll}
\\
\hline
 & Beam (mas) & rms ($\mu$Jy\,beam$^{-1}$) & Peak (mJy\,beam$^{-1}$) & Integrated (mJy\,beam$^{-1}$)\\
\hline
L-band\\
Natural & 13.06 $\times$ 5.01 & 18 $\pm$ 2& 31.40 $\pm$ 3 & 36.8 $\pm$ 4\\
Uniform & 9.79 $\times$ 3.40 & 40 $\pm$ 4 & 29.03 $\pm$ 3 & 32.3 $\pm$ 3\\
\hline
S-band\\
Natural & 8.18 $\times$ 3.42 & 41 $\pm$ 4 & 38.68 $\pm$ 4 & 47.2 $\pm$ 5\\
Uniform & 5.75 $\times$ 2.22 & 62 $\pm$ 6 & 27.80 $\pm$ 3 & 36.0 $\pm$ 4\\
\hline
X-band\\
Natural & 2.13 $\times$ 0.86 & 14 $\pm$ 1 & 61.01 $\pm$ 6 & 68.4 $\pm$ 7\\
Uniform & 1.59 $\times$ 0.53 & 29 $\pm$ 3 & 54.5 $\pm$ 5 & 66.0 $\pm$ 7\\
\hline
\end{tabular}
\end{center}

\end{table*}

\section{Results}

\begin{figure}
\includegraphics[angle=-90, scale=0.3]{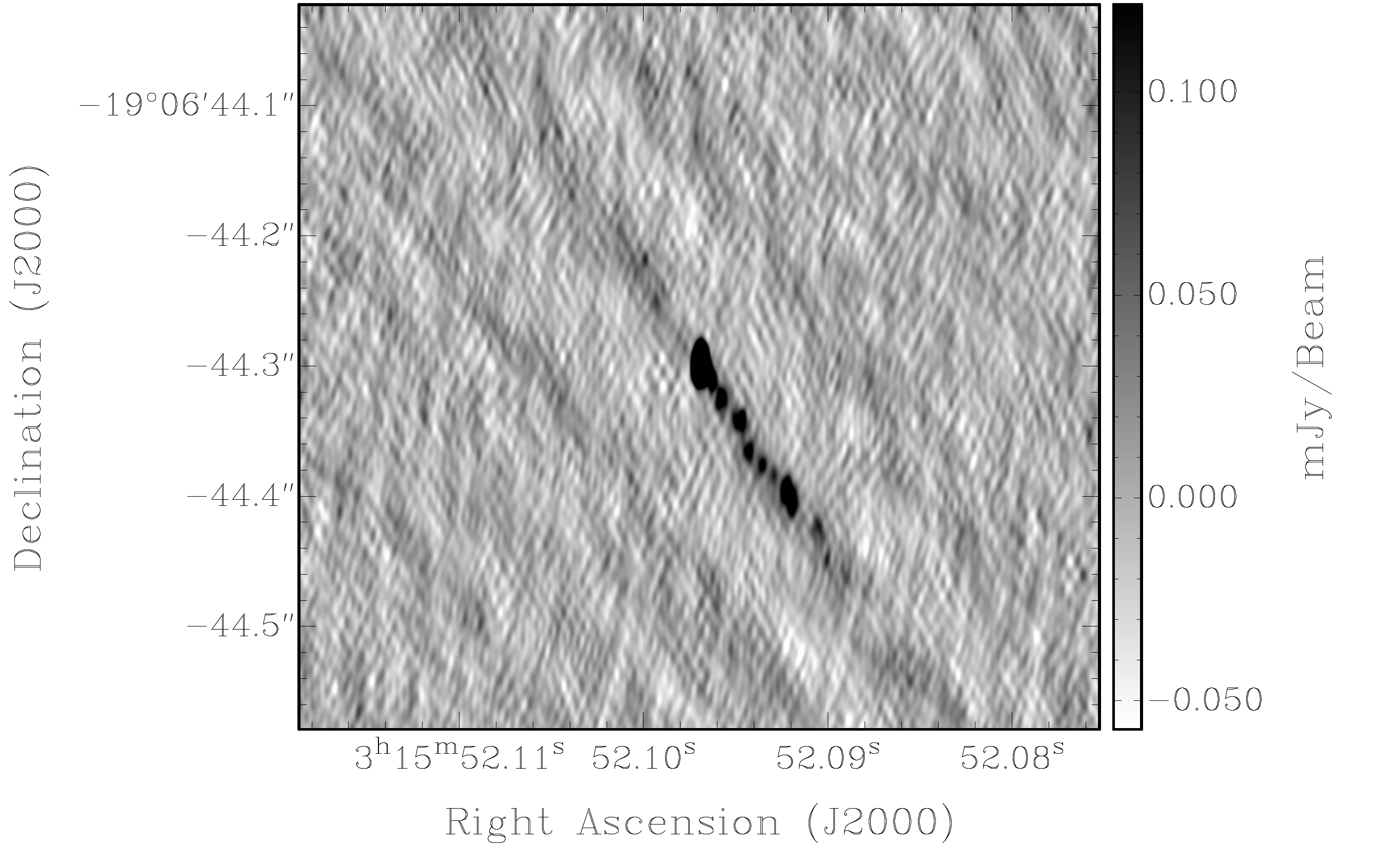}\\
\includegraphics[angle=-90, scale=0.3]{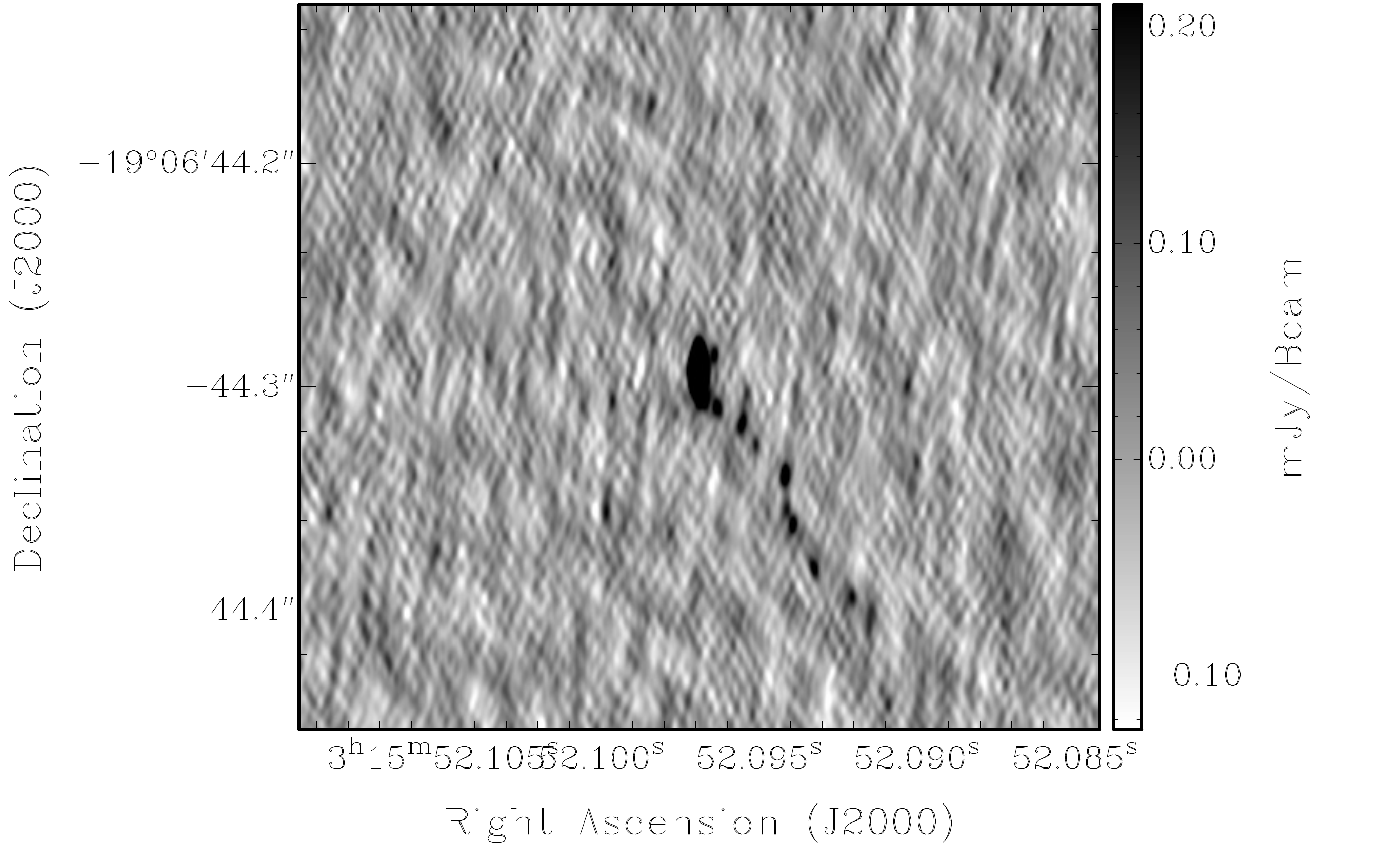}\\
\includegraphics[angle=-90, scale=0.3]{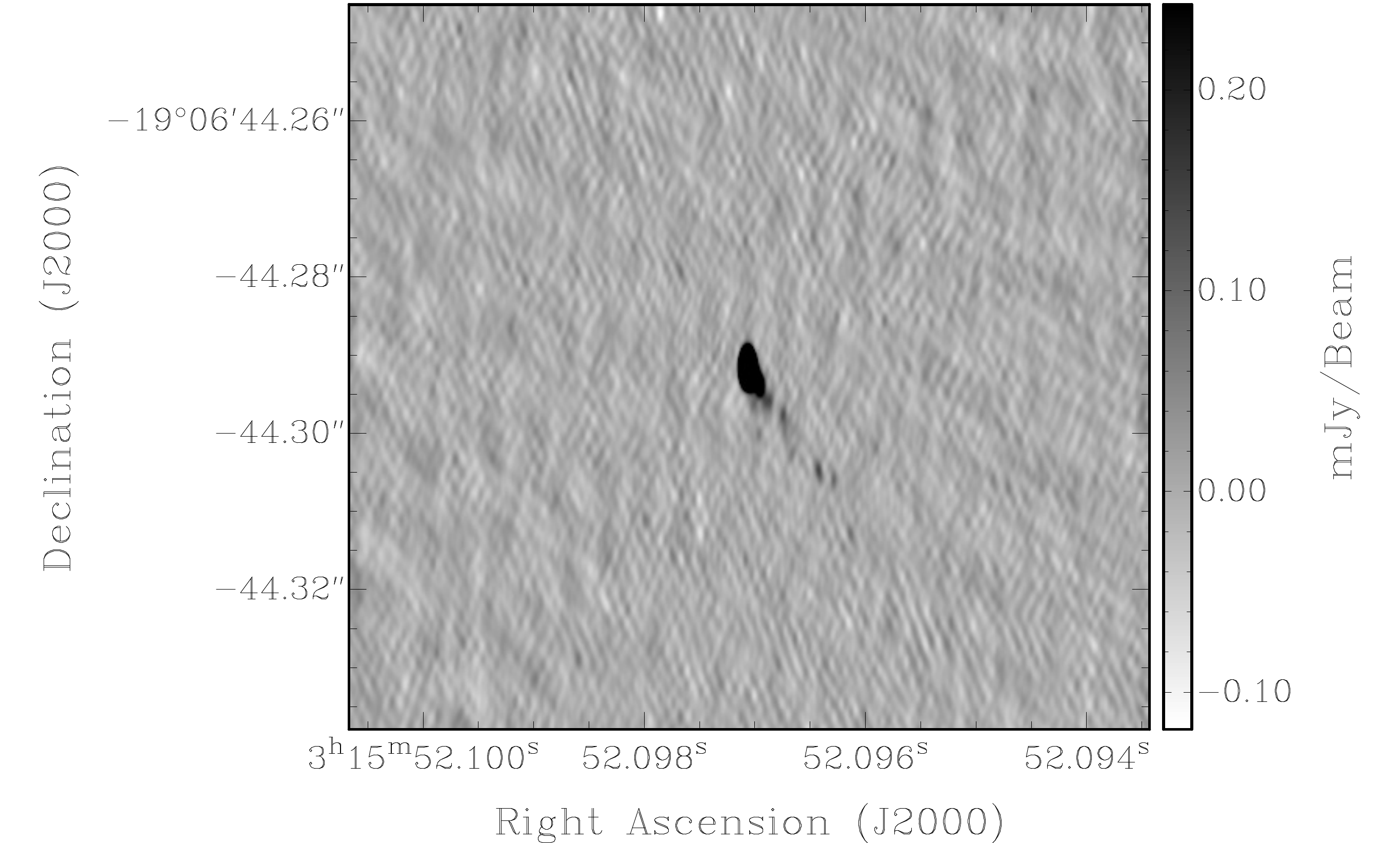}\\
\caption{Naturally-weighted grey-scale images of 0313$-$192 at: \textbf{Top:} L-band \textbf{Middle:} S-Band \textbf{Bottom:} X-Band. }\label{image}
\end{figure}

The core of 0313$-$192 is detected at S-band with the LBA, and L-, S-, and X-band with the VLBA. The LBA observations detected an unresolved point source with a flux density of $53.6 \pm 8.4$\,mJy. The longer baselines of the VLBA enabled higher resolution images that were able, at all three observed frequencies, to resolve the radio structure. Table \ref{flux} provides a summary of the VLBA data.

\subsection{VLBI detection and Chance-alignment}
\citet{Keel06} were able to constrain the optical and radio positions to within 0.2\,$''$. The X-band VLBA data provide the best positional information due to the smaller beam size. The error in the radio astrometry is dominated by phase calibration errors, resulting in a positional accuracy of approximately 0.48 mas. As stated in \citet{Keel06}, the coincidence of the VLBI detections of a compact source with the optical centre of 0313$-$192 effectively eliminates any possibility of the spiral host and DRAGN being in a chance-alignment. A $\sim$30\,mJy compact source detected at L-band using the VLBA has a brightness temperature of T$_{b}>10^{8}$, which can only be attributed to the presence of an AGN. The integrated flux of the naturally weighted S-band source agrees within the errors with the S-band LBA observations from 2009 suggesting little variability on this timescale. The compact radio detection at the position of the optical host, and the detection of jet components, confirms the presence of an active AGN. 



\subsection{VLBA-detected Core and Jet}
\citet{Ledlow01} report an inverted spectral index\footnote{where  $S \propto \nu^{\alpha}$} of the radio core of 0313$-$192 with $\alpha^{4.9 \textrm{GHz}}_{43 \textrm{GHz}} = 0.2$, which is common for AGN cores. They also note the steeper inverted spectrum between 1.4\,GHz and 4.9\,GHz, likely due to either free-free or synchrotron self absorption. The new VLBA observations confirm the inverted nature of the spectral index of the core. We measure a core spectral index of $\alpha^{4.9 \textrm{GHz}}_{43\textrm{GHz}} = 0.37 \pm 0.05$ using the L, S, and X-band VLBA measurements of the unresolved radio core. 

Previously published VLA data of 0313$-$192 at L- and X-band suggest the presence of a radio jet to the South of the host galaxy \citep{Ledlow01}. Although the diffuse radio lobes are detected on either side of the host galaxy, the putative jet suggests that the radio jets are oriented with the Southern jet emitting towards the line-of-sight. 

We detect components of the Southern jet at L-, S-, and X-band, using the VLBA. Although there are tentative detections of components in the Northern jet at S-band, these are not detected at L- and X-bands suggesting either components of highly unusual spectral indices, or more likely, imaging artefacts at S-band. The S-band observations were considerably plagued with RFI, requiring flagging of  $>$50 percent of the data (5 out of 8 sub-bands were flagged almost completely, in addition to time based flagging per station). Some of the remaining S-band data are likely still affected by the RFI resulting in amplitude errors which tend to produce symmetrical artifacts \citep{Walker1995}. The L-band data also required $\sim$25 percent of the data to be flagged (2 out of 8 sub-bands were flagged almost completely). The detection of the Southern jet with the VLBA confirms the hypothesis that this radio source is oriented with the Southern jet coming towards the line-of-sight. 

\begin{figure}
\begin{center}
\includegraphics[angle=0, scale=0.33]{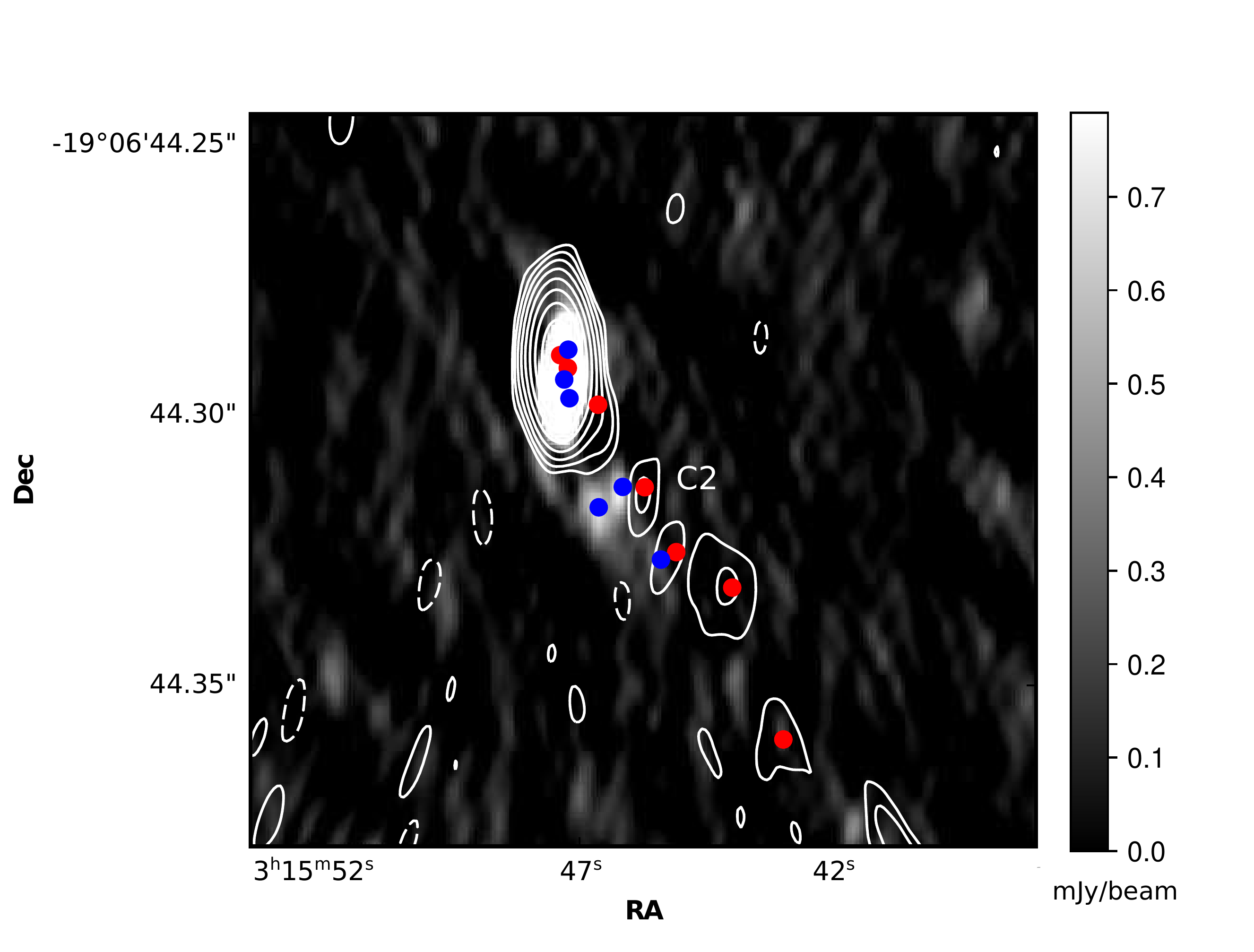}
\end{center}
\caption{S-band modelfit image (grayscale) with L-band modelfit contours (white) overlaid with the S-band (blue) and L-band (red) component positions. Note component size is not shown in order to aid viewing. All components were modeled as circular gaussians using Difmap. The component marked C2 in the image was used for jet-counter jet ratio calculations, being the first discrete component visible in both the L-band and S-band images. }\label{fig:s-l-comps}

\end{figure}
Assuming a symmetrical jet as described in \citet{Lind85}, the jet-to-counterjet ratio is given by:

\begin{equation}
\frac{S_a}{S_r} = \left(\frac{1 + \beta \cos\theta}{1 - \beta \cos\theta}\right)^{n-\alpha}
\end{equation} Where $S_a$ is the flux density of a component in the approaching jet, $S_r$ is the flux density of a component in the receding jet, index $n=2$ for an unresolved `flow' or $n=3$ for a resolved `knot', $\alpha$ is the spectral index of the component, $\beta$ is the jet speed ($\beta=v/c$) and $\theta$ is the inclination angle of the source to the line of sight. 

Using component C2 shown in Figure \ref{fig:s-l-comps}, which has a spectral index $\alpha = -0.333$ we find $S_a = 456$ $\mu$Jy at L-band. The absence of a clear counterjet detection 
puts an upper limit on $S_r < 3\sigma = 54$ $\mu$Jy. Since C2 is resolved we use $n=3$. For the limit case of $\beta=1$ this gives an upper limit to the inclination angle of $\theta \lesssim 72$ degrees. For a less relativistic jet $\beta=0.5$, $\theta \lesssim 52$ degrees.  

At S-band the same calculation with $S_a = 381$ $\mu$Jy, $S_r < 3\sigma = 123$ $\mu$Jy and $\beta=1$ gives a limit of $\theta \lesssim 80$ degrees. At X-band, $S_a = 273$ $\mu$Jy, $S_r < 3\sigma = 42$ $\mu$Jy and $\beta=1$ gives a limit of $\theta \lesssim 74$ degrees.


These values are in good agreement with those calculated by \cite{Keel06}, who found $\theta \lesssim 75$ degrees using X-band VLA data as well as Chandra X-ray data. This also remains consistent with a model of the source in which the jets lie relatively close to perpendicular to the host galaxy and so interact with a minimum of interstellar gas, further evidenced by their linear nature.

The clear detection of a jet with discrete components provides (particularly at X-band) a first epoch for proper motion studies of the jet. Further epochs would allow determination of a (possibly superluminal) jet speed which would give a tighter limit to the inclination angle.

The stellar density in the centre 100\,pc of the Milky Way is of order $10^3$ stars/pc$^{3}$ \citep{Oort71}. As can be seen in Figure \ref{image}, the pc scale jet extends out at least 250 mas which at a redshift of 0.067 gives a physical scale of approximately 300 pc. Considering an `average' jet width of order 50 pc \citep{Helmboldt2008}, and assuming a similar stellar core density in 0319-192 as in the Milky Way, 
this gives a high likelihood of jet-star interaction as proposed by \cite{Huarte-Espinosa13}. No clear evidence of such interaction is seen in the VLBI images presented here, however further epochs of observation would allow for the identification of stationary components/features indicative of jet-star interactions as seen in, for example, Centaurus A \citep{Mueller14}.

\begin{figure*}
\begin{center}
\includegraphics[angle=0, scale=0.32]{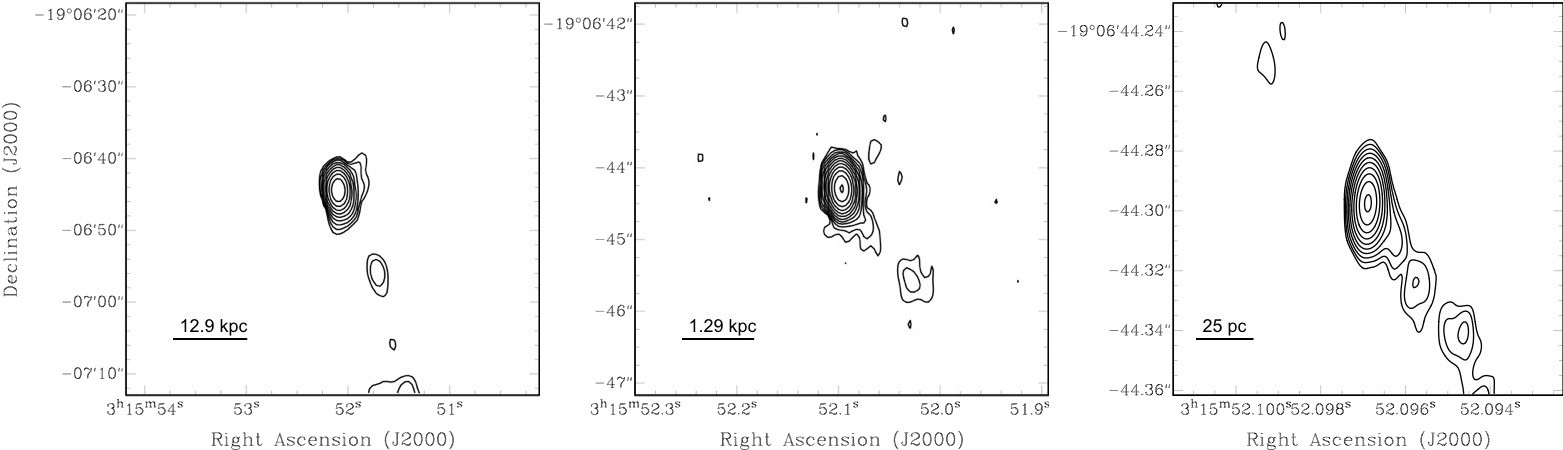}\\
\end{center}
\caption{The radio structure of 0313$-$192 at various scales highlighting the constant jet angle from L-band VLA (\textbf{left}), to X-band VLA (\textbf{centre}), to L-band VLBA (\textbf{right}). Contours are shown at 3$\sigma$ and increase by factors of 2. }\label{zoomfig}
\end{figure*}

The presence of a DRAGN within a spiral host suggested that we may detect interaction of the radio jet with the ISM of the host galaxy at VLBI resolutions. Curiously, the detected components of the Southern jet appear to be in excellent alignment ($\lesssim$10\,degrees) over pc to kpc scales (Figure \ref{zoomfig}). All three observed VLBA frequencies detect radio components of the Southern jet at the same angle to the core, and this angle is seen in the lower-resolution X-band VLA observation, as well as the even lower resolution L-band VLA observation. The L-band VLA observations from \citet{Ledlow01} do appear to show a slight change in angle of the lobe outside the scale-height of the host galaxy. An explanation for this apparent lack of interaction with the spiral host galaxy's ISM may be due to the orientation of the radio structure with respect to the disk of the host galaxy. As the jets/lobes are largely aligned with the minor axis of the host galaxy, perhaps there is simply less ISM for the radio jets to interact with. The large-scale radio lobes indicate sustained and/or repeated phases of AGN activity over timescales of $>10^7$\,years. It is possible that the radio jets from earlier epochs have cleared out the ISM on this trajectory.

\begin{figure}
\begin{center}
\hspace*{-1cm}
\includegraphics[scale=0.215]{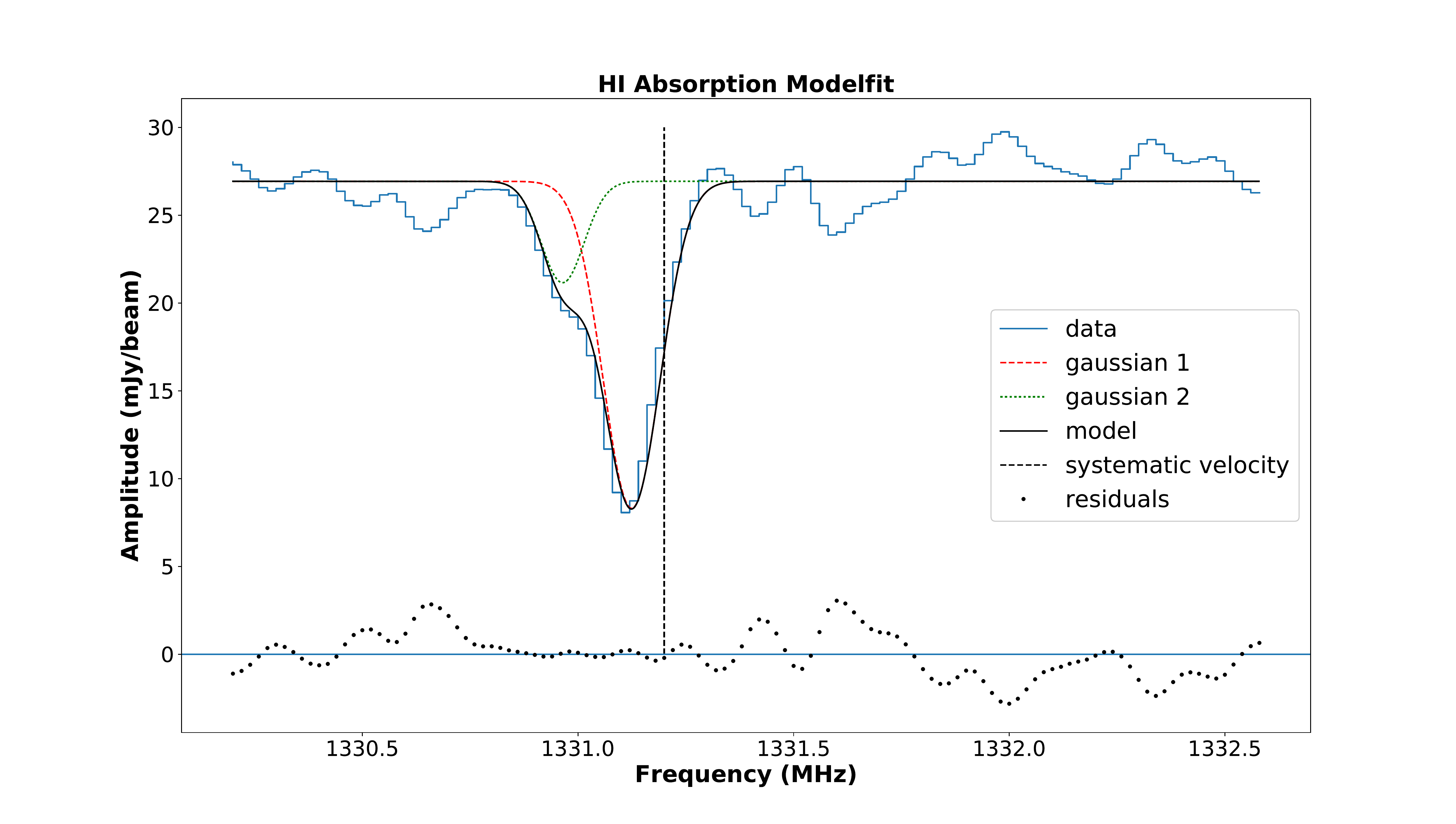}\\
\caption{HI spectrum for 0313$-$192 fit with a two component (red and green dotted lines) gaussian model (black line). The frequency of the systematic velocity ($V_{helio}=20100$ km/s) is shown as the vertical dashed line. Residuals are plotted below (black points).}\label{HIspec-fit}
\end{center}
\end{figure}


\subsection{HI}
\citet{Ledlow01} find HI absorption at $\nu=20,106$ km s$^{-1}$ with optical depth $\tau = 0.98$ and a narrow line width (FWHM) of $\sim 34$ km s$^{-1}$. The first sub-band of the VLBA observations presented in this paper was centred at the redshifted HI frequency of 1331.214\,MHz ($v \sim 20100$ km s$^{-1}$, and the central 4 MHz correlated with a spectral resolution of 20 kHz (4.5 km $s^{-1}$). The resulting spectrum is shown in Figure \ref{HIspec-fit} and is in good agreement with that of \citet{Ledlow01}. 

The absorption is best fit by a two component Gaussian model: The main component is at a frequency of 1331.124 MHz ($v \sim 20122$ km s$^{-1}$), with FWHM $0.134 \pm 0.009$ MHz ($\sim 30.2$ km s$^{-1}$) with the secondary component at frequency 1330.967 MHz ($v \sim 20160$ km s$^{-1}$) with FWHM $0.0914 \pm 0.02$ MHz ($\sim 22$ km $s^{-1}$). Evidence for this second component is visible in the data presented by \citet{Ledlow01} (but is not discussed), but is confirmed by the higher-resolution VLBA spectrum, perhaps suggesting the presence of an outflow (or inflow). These components lie at velocities of 22 km/s and 60 km/s respectively from the systematic velocity of $V_{helio} = 20100$ km/s. As noted by \citet{Ledlow01}, the majority of HI absorption features seen in AGN hosts show line widths in excess of 50 km/s, produced by features close to the nucleus (and hence with a high range of velocities). In this case it is thus likely that the absorption is originating from gas in the ISM of the galaxy, perhaps in a spiral arm or dense cloud.


No perturbed neutral gas is detected spatially, but this may be due to the lack of continuum emission detected away from the nucleus, making it difficult to probe the gas kinematics on larger scales.


\section{Conclusions}
This paper presents the first observations of a spiral DRAGN at VLBI resolutions. We have detected the core of 0313$-$192, the archetypal spiral DRAGN, using both the LBA and the VLBA. The longer baselines of the VLBA enabled the LBA-detected point-source to be resolved. The detection of the radio core of 0313$-$192 eliminates the chance that the DRAGN was merely chance aligned with the spiral host. Moreover, detecting the core at VLBI resolutions confirms the presence of an active AGN in the core of this spiral galaxy. The approaching jet is detected at all three observed frequencies of the VLBA. Assuming a symmetric, relativistic jet we were able to estimate an upper limit to the inclination angle of $\theta \lesssim 72$, which is the best constraint made to-date. Perhaps unexpectedly, the VLBI detected radio jet is in excellent alignment with the larger-scale jet components detected with the VLA at L- and X-band. This suggests little if any disruption to the jet by the host galaxy's ISM. We suggest this may be due to the orientation of the DRAGN as it is aligned with the minor axis of the galaxy. 

Currently accepted galaxy formation models require a major merger, where both galaxies are of similar size, to trigger a DRAGN. Spiral galaxies are not expected to withstand a major merger. Consequently, the existence of spiral DRAGNs challenges these models. Our results show that, even on VLBI-scales, the radio jet is being emitted perpendicular to the stellar disk of the spiral galaxy. We see no evidence for interaction by the jets with the spiral galaxy's ISM. The DRAGN hosted by 0313$-$192 has a total spatial extent of 360\,kpc. The large-scale nature of the radio lobes may suggested sustained or repeated phases of AGN activity over timescales of $>$10$^7$ years. A possible explanation for linear nature of the VLBI-scale jet may be that the radio jets from earlier epochs have already cleared out material in their path. 

As suggested by \citet{Keel06}, perhaps a minor merger or interaction was sustained by 0313$-$192 earlier in its lifetime that triggered the DRAGN, but was sufficiently minor to not disturb the spiral host completely, as seen in some Seyfert galaxies \citep[e.g.][]{Taniguchi99}. The serendipitous orientation of the jet to the stellar disk and the lack of disturbance to the radio structure over the last $>$10$^7$ years, as evidenced by the linear structure on pc to kpc scales, may then explain the prevalence, and hence size, of the DRAGN associated with 0313$-$192.

All of the currently detected spiral DRAGNs reside in galaxy groups or clusters, show evidence for prior mergers and/or interactions, contain prominent bulges, and have their large-scale radio lobes perpendicular to the stellar disk of the spiral host. 

Spiral DRAGNs appear to require very specific conditions in order to exist: a minor merger earlier in their lifecycle that doesn't disturb the spiral structure too greatly, a denser-than-normal environment, and jets that are perpendicular to the stellar disk.

Although more spiral DRAGNs must be investigated in order for these parameters to be quantified, the evidence to-date suggests that DRAGNs may perhaps require mergers that were more minor than previously hypothesized. However, the small number of detected spiral DRAGNs in the local Universe attests to the specificity of the conditions required to produce these enigmatic objects. 

\section*{Acknowledgements}
MYM acknowledges support from EC H2020-MSCA-IF-2014 660432 (spiral DRAGNs), and also from NRAO for providing a workspace and resources. This paper includes data from the VLBA, which is operated by the Long Baseline Observatory (LBO). The LBO is a facility of the National Science Foundation operated under cooperative agreement by Associated Universities, Inc. The Australia Telescope Compact Array/Long Baseline Array is part of the Australia Telescope National Facility which is funded by the Australian Government for operation as a National Facility managed by CSIRO. This work made use of the Swinburne University of Technology software correlator, developed as part of the Australian Major National Research Facilities Programme. We are grateful for the many useful comments from the MNRAS Science Editor and the insightful suggestions made by our wonderful referee, Professor Eric S. Perlman.












\bsp	
\label{lastpage}
\end{document}